\documentclass[aps,prd,amsfonts,amsmath,twocolumn,nofootinbib]{revtex4}

\usepackage{color}
\usepackage{graphicx}

\usepackage{epstopdf}
\newcommand {\be} {\begin{equation}}
\newcommand {\ba} {\begin{eqnarray}}
\newcommand {\ee} {\end{equation}}
\newcommand {\ea} {\end{eqnarray}}

\begin{document}
\title{Strange mesons and kaon-to-pion transition form factors from holography}
\author{Zainul Abidin and Carl E.\ Carlson}
\affiliation{
Department of Physics, College of William and Mary, Williamsburg, VA 23187, USA}

\date{August 17, 2009}

\begin{abstract}
We present a calculation of the $K_{\ell 3}$ transition form factors using the AdS/QCD correspondence.  We also solidify and extend our ability to calculate quantities in the flavor-broken versions of AdS/QCD.  The normalization of the form factors is a crucial ingredient for extracting $|V_{us}|$ from data, and the results obtained here agree well with results from chiral perturbation theory and lattice gauge theory.  The slopes and curvature of the form factors agree well with the data, and with what results are available from other methods of calculation.
\end{abstract}

\maketitle

%%%%%%%%%%%%%%%%%%%%%%%%%%%%%%%%%%%%%%%%%%%%%%%
%
\section{Introduction}
%
%%%%%%%%%%%%%%%%%%%%%%%%%%%%%%%%%%%%%%%%%%%%%%%
In this paper, we consider the extension of the anti-de Sitter space/quantum chromodynamics model (AdS/QCD) to allow broken flavor symmetry, and apply the model to the kaon system and particularly to the $K_{\ell 3}$ form factors.

The connection between 5D gravitational theories on an anti-de Sitter space and 4D conformal field theories began as a correspondence between a type IIB string theory and an $\mathcal{N}=4$ super Yang-Mills theory in the large $N_C$ limit~\cite{Maldacena:1997re,Witten:1998qj,Gubser:1998bc}.  This has inspired an analytic model referred to as AdS/QCD connecting 5D theories living on an anti-de Sitter space to 4D QCD~\cite{Erlich:2005qh,DaRold:2005zs}.  Interesting results have been obtained for masses, couplings, and electromagnetic and gravitational form factors for vector, axial vector and pseudoscalar mesons.  For a selection of these results, see~\cite{Polchinski:2001tt,Polchinski:2002jw,Brodsky:2003px,deTeramond:2005su,Brodsky:2006uqa,Brodsky:2008pf,Grigoryan:2007vg,Grigoryan:2007my,Grigoryan:2007wn,Grigoryan:2008cc,Kwee:2007dd,Kwee:2007nq,BoschiFilho:2005yh,Abidin:2008ku,Abidin:2008hn,Abidin:2008sb,Abidin:2009hr}.
%
%{\vglue -81.0 mm
%\hglue -16 mm
%\includegraphics{text2451v101.eps}
%\vglue -46.7 mm
%}

The $K_{\ell 3}$ form factors describe the decays $K \to \pi \ell \nu$ and are  currently most known for the role they play in high-precision extractions of the Cabibbo-Kobayashi-Maskawa (CKM) matrix element $V_{us}$.  The form factors themselves are the $K$ to $\pi$ transition matrix elements of the strangeness changing vector current.  There are two $K_{\ell 3}$ form factors, bespeaking the fact that the strangeness changing current is not conserved, and has a longitudinal as well as a transverse part.  Experiments measure the product of the form factors and $|V_{us}|$.  Hence the extraction of $|V_{us}|$ from the $K_{\ell 3}$-decay measurements depends on having a reliable calculation of the form factor normalization.  So far, these calculations have been from chiral perturbation theory~\cite{Leutwyler:1984je,Bijnens:2003uy,Cirigliano:2005xn,Jamin:2006tj} or from lattice gauge theory~\cite{Becirevic:2004ya,Dawson:2006qc,Boyle:2007qe,Flynn:2008hd,Lubicz:2009ht}.  Here we will present a first calculation of these form factors from AdS/QCD.  We will also solidify and extend our ability to calculate quantities in the flavor-broken versions of AdS/QCD.

Our form factor normalizations can be compared to those obtained from other calculational methods, our slopes can be compared to data as well as to other calculations, and since we calculate using an analytic method, we can also obtain a curvature that can be compared to experimental data.  All comparisons of results to other methods and to data show good agreement, as will be detailed below.

We work with general mass pseudoscalar mesons.  Previous results known to us worked in the chiral limit or calculated some quantities using expansions valid at small mass.  In particular, while we can neatly derive the Gell-Mann-Oakes-Renner (GOR)~\cite{GellMann:1968rz} relation in the chiral limit, we do not use it to obtain any of our results and can test to see how accurate it is at given quark mass.

Section~\ref{sec:adsqcd} reviews AdS/QCD with notation pertinent to several quarks of differing masses. Section~\ref{sec:twopoint} gives results obtained from two-point functions, including bulk-to-boundary propagators, masses, and decay constants, focusing on differences from the flavor symmetric case, in particular considering scalar mesons along with vector, axial vector, and pseudoscalar mesons.  Section~\ref{sec:threepoint} contains algebraic results for the form factors, with the numerical results given in Sec.~\ref{sec:numerical}.  Closing remarks are offered in Sec.~\ref{sec:end}.

%%%%%%%%%%%%%%%%%%%%%%%%%%%%%%%%%%%%%%%%%%%%%%%
%
\section{The AdS/QCD model} \label{sec:adsqcd}
%
%%%%%%%%%%%%%%%%%%%%%%%%%%%%%%%%%%%%%%%%%%%%%%%

We will use the following metric for the 5 dimensional Anti-de Sitter space 
\be
ds^2=\frac{1}{z^2}\left(\eta_{\mu\nu}dx^\mu dx^\nu-dz^2\right),\qquad \varepsilon<z<z_0,
\ee 
where the metric of the 4 dimensional flat space is $\eta_{\mu\nu}=$diag$(1,-1,-1,-1)$. The cut-off at $z=\varepsilon$ (with $\varepsilon \to 0$ implied) corresponds to UV cut-off in QCD, while the hard-wall cut-off at $z_0$ corresponds to IR cut-off, $\Lambda_{QCD}$, to simulate confinement. 

According to the AdS/CFT correspondence, for  every operator in 4 dimensional theory there is a corresponding  field in the AdS$_5$ space.  Operators of our interest are current operators ${J^a_L}_\mu=\bar{\psi_q}_L \gamma_\mu t^a{\psi_q}_L$, ${J^a_R}_\mu=\bar{\psi_q}_R \gamma_\mu t^a{\psi_q}_R$ and quark bilinear $\bar {\psi_q}_L {\psi_q}_R$. In the AdS$_5$ space, these operators correspond to gauge fields $L^a_\mu(x,z)$, $R^a_\mu(x,z)$,  and  a scalar field $X(x,z)$ respectively. Following \cite{Erlich:2005qh}, we will consider a 5D action with SU(3)$_L \otimes$\,SU(3)$_R$ symmetry as follows
\be
S = \int d^5 x \, \sqrt{g} \, {\rm Tr}\Big\{\left|D X\right|^2+3\left|X\right|^2-\frac{1}{4g_5^2}\left(F_{(L)}^2+F_{(R)}^2\right)\Big\}  .
\ee
The field strength is defined by $F^{(L)}_{MN}=\partial_M L_N-\partial_N L_M-i[L_M,L_N]$ and analogously for $F^{(R)}_{MN}$. The scalar field $X$ and gauge fields interact through the covariant derivative $D_M X=\partial_M X -iL_M X +i X R_M$ in such a way that the action is chiral invariant.   We also use the vector and the axial-vector field defined from $L=V+A$ and $R=V-A$.

The theory begins as one that has SU(3)$_L \otimes$\,SU(3)$_R$ symmetry, and one would like to maintain as much of the symmetry as possible even when going to massive quarks and in particular to flavor non-symmetric quark masses.  In a chirally symmetric world, the action is invariant as the $X$ field transforms via
\be
X \to X' = U_L X U_R^\dagger		.
\ee
One can expand X as
\be
X(x,z) = e^{i \pi^a(x,z) t^a}  X_0(z) e^{i \pi^a(x,z) t^a}
\ee
whereupon an axial transformation (which has $U_L^\dagger = U_R$) induces to leading order a shift in the pion field,
${\pi'}^a = \pi^a - \theta^a$, where $\theta^a$ is a parameter in the transformation $U_L = e^{-i \theta^a t^a}$ and is consistent with $\pi^a$ being a pseudoscalar field.  With flavor symmetry, $X_0$ is a multiple of the unit matrix and commuting, so one can easily write $X = e^{2i \pi^a t^a}  X_0$, as has often been done.  However, in the flavor non-symmetric case, this would make $\pi^a$ appear to be associated with left-handed transformations rather than with axial transformations, and gives it unexpected parity properties and mixing with vector as well as with axial vector fields.  For example, one obtains a quadratic term in the Lagrangian proportional to
$\eta^{MN} {\rm Tr}\, [X_0,\partial_M \pi^b t^b] [X_0, V_N]$, which will violate 4D parity when $X_0$ is not proportional to the unit matrix.  (With the split exponential, one gets cancellations rather than just commutators.)

Shock and Wu~\cite{Shock:2006qy} have early on considered three-flavor extensions of AdS/QCD, although keeping $X = e^{2i \pi^a t^a}  X_0$.  They did not study the more dynamical quantities like form factors, but did obtain many good results for masses and decay constants.  However, as they themselves point out, they did not obtain good results for the ground state pseudoscalar kaons with the same parameters that gave good results for the more excited strange mesons.  With the exponential split, as we think it should be, one obtains good results for pseudoscalar as well as strange axial and vector (and even scalar) meson states.

Katz and Schwartz~\cite{Katz:2007tf} also considered flavor-broken AdS/QCD, although their main focus was on the U(1) problem and also did not study form factors.  They also kept 
$X = e^{2i \pi^a t^a}  X_0$, but only explicitly studied the part of the action that mixes the axial vectors with the pseudoscalars, where problems do not appear.  We might remark already that they used the GOR relation to get the strange quark mass.  The GOR relation becomes less valid as the quark mass increases, and using a different method to fix the strange quark mass, we find a larger value than they quote.

Hambye \textit{et al.}~\cite{Hambye:2006av} also studied three-flavor AdS/QCD, focusing on quantities that are calculated from four-point functions such as the purely hadronic $K_{\pi 2}$ decays or the $B_K$ parameter needed to calculate $K^0$-$\bar K^0$ mixing.  They work in a limit where all quarks are massless, and so have $X_0 = X = 0$.  Hence their subjects and their approximations do not greatly overlap with the present paper, although we plan to consider quantities obtained from four-point functions in future work.

Turning off all fields  except $X_0(z)$ and solving the equation of motion, one obtains   
\be
2{X_0}_{ij}=v_{ij}(z)= \zeta M_{ij} z+ 
	\frac{1}{\zeta} \Sigma_{ij} z^3\,, \label{vacuumexp}
\ee
where $\zeta$ is a rescaling parameter~\cite{DaRold:2005vr,Cherman:2008eh} discussed below.  From the AdS/CFT correspondence $M_{ij}$ can be identified as quark mass matrix which responsible for the explicit breaking of the chiral symmetry of QCD and $\Sigma_{ij}$  as the quark condensate $\left<\bar q_i q_j\right>$ which spontaneously break the chiral symmetry of QCD to SU(3)$_V$. Assuming $u$ and $d$ symmetry, we have
\be
M=\left(\begin{array}{ccc}
                  m_q &   0      & 0\\
                    0     & m_q  & 0\\
                    0     &  0       & m_s
                  \end{array}
	\right)\,,\qquad 
\Sigma=\left(\begin{array}{ccc}
                  \sigma_q &   0      & 0\\
                    0     &  \sigma_q  & 0\\
                    0     &  0       &  \sigma_s
                  \end{array}
	\right)\,.
\ee 
In general $\sigma_s \neq \sigma_q$. However,  we will also consider the limiting case when  $\sigma_s =\sigma_q$, as an analytic solution for the vector field can be obtained.

Regarding the quark masses and the quark condensate parameter $\sigma$, we adopt a normalization parameter as advocated in~\cite{DaRold:2005vr,Cherman:2008eh}, wherein quark masses are multiplied by a factor $\zeta = \sqrt{N_C}/2\pi$~\cite{Kapusta:2009ab} compared to earlier conventions and $\sigma$ is divided by the same factor.  One can, of course, view the quark masses and $\sigma$ in AdS/QCD as parameters of this particular model, and many important quantities, including the GOR relation and most of the results in this paper, are unchanged by this rescaling.   However, the rescaled parameters allow a precise connection to the two-point correlation function of the quark condensate at small distances, which is known from perturbative QCD, and also leads to a better agreement with mass parameters at the hadronic scale and with the quark condensate as obtained from methods disconnected from AdS/QCD.  Quark masses obtained from AdS/QCD had generally been strikingly low and the quark condensate strikingly high, but following~\cite{DaRold:2005vr,Cherman:2008eh} one can argue that the disagreement was a matter of having incommensurate definitions.

%%%%%%%%%%%%%%%%%%%%%%%%%%%%%%%%%%%%%%%%%%%%%%%%
%
\section{Two-point function} \label{sec:twopoint}
%
%%%%%%%%%%%%%%%%%%%%%%%%%%%%%%%%%%%%%%%%%%%%%%%%%%

Up to second order in fields, the action can be written as
\begin{align}
S&=\int\, d^5 x  \bigg\{\sum_a \frac{-1}{4g_5^2 z} (\partial_M V^a_N-\partial_N V^a_M)^2+\frac{ {M_V^a}^2(z)}{2z^3}{V^a_M}^2\nonumber\\ 
&\frac{-1}{4g_5^2 z} (\partial_M A^a_N-\partial_N A^a_M)^2+\frac{ {M_A^a}^2(z)}{2z^3}(\partial_M\pi^a-A^a_M)^2 \bigg\}, \label{action2nd}
\end{align}
where contraction over $\eta_{ML}$ is implicit.   The mass combinations come from
\ba
\frac{1}{2} {M_V^a}^2 \delta^{ab} &=& 
			- {\rm Tr\,} \left[t^a,X_0 \right] \left[t^b,X_0 \right] \,,
					\nonumber \\
\frac{1}{2} {M_A^a}^2 \delta^{ab} 
			&=&  {\rm Tr\,} \left\{t^a,X_0 \right\} \left\{t^b,X_0 \right\}	\,,
\ea
or,
\ba
{M_V^a}^2 &=& \left \{ 
\begin{array}{ll} 
0 & a=1,2,3\\ 
\frac{1}{4}\left(v_s-v_q\right)^2 \quad &  a=4,5,6,7\\ 
0 &  a=8 \,,
\end{array} \right .\nonumber\\
{M_A^a}^2   &=&   \left \{ 
\begin{array}{ll} 
v_q^2 & a=1,2,3\\ 
\frac{1}{4}\left(v_q+v_s\right)^2 \quad  &  a=4,5,6,7\\ 
\frac{1}{3}\left(v_q^2+2v_s^2\right) &  a=8      \,,
\end{array} \right .\nonumber\\
\ea
where 
\ba
v_q(z) &=& \zeta m_q z + \frac{\sigma_q}{\zeta} z^3   \,,    \nonumber \\
v_s (z)&=& \zeta m_s z + \frac{\sigma_s}{\zeta} z^3  \,.
\ea

\noindent   For later convenience we define
\ba
\alpha^a(z) = \frac{g_5^2 {M_V^a}^2}{z^2} \,,   \quad
\beta^a(z) = \frac{g_5^2 {M_A^a}^2}{z^2}  \,.
\label{eq:alphafunction}
\ea

\noindent As shown in Eq.~\eqref{vacuumexp}, the vacuum solution contains both explicit and spontaneous  symmetry breaking parameters, $M_{ij}$ and $\Sigma_{ij}$ respectively. The parameters $m_q$ and $m_s$ in the 5D theory are usually considered to be explicit symmetry breaking, and give quark mass terms in the 4D theory that are also explicit symmetry breaking.  The condensate parameters may be considered spontaneous symmetry breaking, but in the absence of the quark mass parameters (\textit{i.e.}, in the chiral limit where the $m_q$ and $m_s$ go to zero), one expects that the condensate parameters are all the same.  Hence one may argue that the differences in the condensate parameters do arise from explicit symmetry breaking.  Since the $M_V^a$ functions depend only on differences $v_s-v_q$, one can say that they would be zero if there were only spontaneous symmetry breaking.  In this limit,  the masses of the vector mesons in the same octet are degenerate. 

The axial sector of action (\ref{action2nd}) is invariant under gauge transformation,
\ba
A^a_M&\rightarrow& A^a_M -\partial_M \lambda^a\,,\nonumber\\
\pi^a&\rightarrow& \pi^a-\lambda^a\,.
\ea
Hence, we are free to set $A^a_z=0$. For the vector sector, the mass term destroys the gauge freedom for $a=4,5,6,7$. Hence, we can set $V^a_z=0$ only for $a=0,1,2,3,8$. We will show that the non-vanishing $V_z$ is related to the non-vanishing longitudinal part of the vector field.

%%%%%%%%%%%%%%%%%%%%%%%%%%%%%%%%%%%%%%%%%%%%%%%%%%
\subsection{Vector sector}
%%%%%%%%%%%%%%%%%%%%%%%%%%%%%%%%%%%%%%%%%%%%%%%%%%

The vector field satisfies the following equation of motion
\be
\eta^{ML}\partial_M\left(\frac{1}{z}\left(\partial_L V^a_N-\partial_N V^a_L\right)\right)+\frac{\alpha^a(z)}{z}V^a_N=0.
\ee
For the transverse part, $\partial^\mu V^a_{\mu\perp}(x,z)=0$,  one obtains
\be
\left(\partial_z\frac{1}{z}\partial_z+\frac{q^2-\alpha^a}{z}\right) V^a_{\mu\perp}(q,z)=0, \label{profiletrans}
\ee
where $q$ is the Fourier variable conjugate to the 4 dimensional coordinates, $x$. 

We shall write the vector field in terms of its boundary value at UV multiplying a profile function, or bulk-to-boundary propagator, $V^a_{\mu\perp}(q,z)= V^{0a}_{\mu\perp}(q) {\cal V}^a(q^2,z)$, and set ${\cal V}^a(q^2,\varepsilon)=1$ (Note that there is no summation over the group index $a$ of the profile function). The boundary value ${V^0_\mu}^a(q)$ acts as the Fourier transform of the source of the 4D conserved vector current operator. At the IR boundary, we choose Neumann boundary condition $\partial_z {\cal V}^a(q,z_0)=0$. In the  $\Sigma_{ij}=\sigma \delta_{ij}$ limit, the solution can be written in terms of Bessel function
\be
{\cal V}^a(q^2,z)=\frac{\pi}{2} \tilde{q} z\left(\frac{Y_0(\tilde{q}z_0)}{J_0(\tilde{q}z_0)}J_1(\tilde{q} z)- Y_1(\tilde{q} z)\right)\,, \label{profileV}
\ee
for $\quad q^2>\alpha^a$, and
\be
{\cal V}^a(q^2,z)=\tilde{Q} z\left(\frac{K_0(\tilde{Q}z_0)}{I_0(\tilde{q}z_0)}I_1(\tilde{Q} z)+ K_1(\tilde{Q} z)\right)\,,
\ee
for $\quad q^2<\alpha^a$, where $\tilde{q}=\sqrt{q^2-\alpha^a}$ and $\tilde{Q}=\sqrt{\alpha^a-q^2}$. Near the UV boundary, the profile function can be written as
\be
{\cal V}(q^2,z)=1+\frac{\tilde q^2z^2}{4} 
	\ln \left(\tilde q^2 z^2\right)+\ldots.
\ee

The longitudinal part of the vector field, $V^a_{\mu\parallel}=\partial_\mu \xi^a$ and $V_z$ are coupled as follows
\begin{align}
-q^2\partial_z \tilde \phi^a(q^2,z) &+\alpha^a\partial_z \tilde \pi^a(q^2,z)=0\,,
				\label{profilelong1}		\\[1.25ex]
\partial_z \bigg(\frac{1}{z}\partial_z\tilde \phi^a(q^2,z) \bigg)  &-\frac{\alpha^a}{z}\big(\tilde \phi^a(q^2,z)-\tilde \pi^a(q^2,z)\big)=0, 
				\label{profilelong2}
\end{align}
where we define $V^a_z=-\partial_z \tilde \pi^a$ and $\xi^a=\tilde \phi^a-\tilde \pi^a$. The constancy of $\alpha^a(z)$ when  $\Sigma_{ij}=\sigma \delta_{ij}$  simplifies above equations into
\be
\left(\partial_z\frac{1}{z}\partial_z+\frac{q^2-\alpha^a}{z}\right)\xi^a(q^2,z)=0.\label{profilelong3}
\ee
This is precisely the equation for the transverse part of the vector field. Fixing boundary conditions as $\tilde \phi^a(q,\varepsilon)=0$ and $\tilde \pi^a(q,\varepsilon)=-1$ on the UV brane and Neumann boundary conditions on the IR brane, one concludes that in the limit where $\sigma_s=\sigma_q=\sigma$ the profile function for the longitudinal and the transverse part of the vector field are identical, $\xi^a(q^2,z)={\cal V}^a(q^2,z)$, with a solution given by Eq.~\eqref{profileV}.  In general, this is not the case, and both $\xi^a$ and ${\cal V}^a$ can be solved numerically. For $a=1,2,3,8$,  longitudinal vector fields are unphysical in the sense that they can be eliminated by fixing the gauge, $V^a_z=0$. 

Two-point functions can be calculated from the AdS/QCD correspondence by evaluating the action (\ref{action2nd}) with the classical solution and taking the functional derivative over $V^0_\mu$  twice. One obtains 
\begin{align}
i\int_x e^{iqx}\left<0\left|\mathcal{T} J^{\mu a}_{\perp}(x)
J^{\nu b}_\perp(0)\right|0\right>&=-P^{\mu\nu}_T \delta^{ab}
\frac{\partial_z {\cal V}^a(q^2,\varepsilon)}{g_5^2 \varepsilon}\,,
		\nonumber\\
i\int_x e^{iqx} \big<  0  \big|\mathcal{T} J^{\mu a}_\parallel(x)
J^{\nu b}_\parallel(0)  \big|  0  \big>
&=-P^{\mu\nu}_L \delta^{ab}
\frac{\partial_z \tilde \phi^a(q^2,\varepsilon)}{g_5^2 \varepsilon}\,, \label{2pointfV}
\end{align}
where $P^{\mu\nu}_T=\left(\eta^{\mu\nu}-q^\mu q^\nu/q^2\right)$ and $P^{\mu\nu}_L=q^\mu q^\nu/q^2$ are the transverse and longitudinal projector respectively. Comparing this result with the quark bubble diagram of QCD,  one can  fix  parameter $g_5$ of the model~\cite{Erlich:2005qh}
\be
g_5^2=\frac{12 \pi^2}{N_c}\,.
\ee

Hadrons correspond to normalizable modes of the 5D fields. These modes should vanish sufficiently fast near the UV brane such that the action is finite and at IR brane satisfy Neumann boundary condition.   The eigenvalue, $q^2=M_n^2$, is the squared mass of the $n$-th Kaluza Klein mode.   We expect vector mesons to be normalizable modes of  equation (\ref{profiletrans}) and scalar mesons to be normalizable modes of  Eqs.\ (\ref{profilelong1}) and (\ref{profilelong2}). In the $\Sigma_{ij}=\sigma \delta_{ij}$ limit,  the scalar meson has identical mass with the corresponding vector meson. However, for $a=1,2,3,8$, longitudinal modes are unphysical. Hence, the lightest scalar meson obtained from the longitudinal mode of the vector field is a strange meson, $K^{*}_0$. Regarding scalar mesons in AdS/QCD, see also Ref.~\cite{DaRold:2005vr}.

As a remark, we may note that one could include a scalar field explicitly by defining $X$, similarly to~\cite{DaRold:2005vr}, as
\be
X = e^{i \pi^a t^a} \left( X_0 + S \right) e^{i \pi^a t^a}
\ee
with $S = S^a t^a$ (ignoring the scalar
singlet).  In this case, after some manipulation, the quadratic terms in the action involving $V^a_M$ and $S^b$ for $a,b = 4,5,6,7$ become
\begin{align}
S&=\int\, d^5 x  \bigg\{\sum_a \frac{-1}{4g_5^2 z} (\partial_M V^a_N-\partial_N V^a_M)^2
    			\nonumber \\
    & \qquad +\frac{ {M_V^a}^2(z)}{2z^3}\left({V^a_M} + \frac{2}{\sqrt{3}}f^{ab8}\partial_M\left(\frac{S^b}{M^a_V}\right)\right)^2\nonumber\\
    &\qquad+\frac{S^a S^a}{2z^5}\left[ 3- z^5\frac{\eta^{MN}}{M^a_V}\partial_M \left(\frac{1}{z^3}\partial_N M^a_V\right) \right] \bigg\}		\,
\end{align}
where one can show that the square bracket on the last line is zero.  Redefining the
the vector field,
\begin{equation} 
V^a_M \rightarrow {V^a_M}
+ \frac{2}{\sqrt{3}}f^{ab8}\partial_M\left(\frac{S^b}{M^a_V}\right) \,,
\end{equation}
one obtains a massive vector field and eliminates the scalar field, and the action becomes like the one we use here.

When the condensate parameters are all the same, wave functions for the vector mesons are given by
\be
\psi^a_n(z)= \frac{\sqrt{2} z J_1(z\sqrt{{M^a_n}^2-\alpha^a})}{z_0 J_1(z_0\sqrt{{M^a_n}^2-\alpha^a})},
\ee
with normalization condition, $\int (dz/z) {\psi^a_n}^2=1$.
In particular, we obtain an infinite tower of KK rho mesons for $a=1,2,3$,  the corresponding tower of  $K^{*}$ mesons for $a=4,5,6,7$, and  $\omega^0$  mesons for $a=8$. The Neumann boundary condition on the IR gives $J_0(M^\rho_n z_0)=0$. Identifying the lightest mode as the rho meson, we can fix $z_0$ parameter of the model.

%%%%%%%%%%%%%%%%%%%%%%%%
\begin{figure}[t]
\begin{center}
%\vglue -10 mm
\includegraphics[width =2.8in]{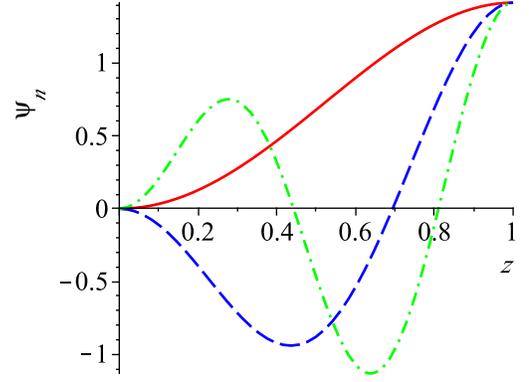}
%\vglue -5 mm
\caption{Plot of $\psi_1$ (solid red curve), $\psi_2$ (dashed blue curve),  and $\psi_3$ (dash-dot green curve), with $z$ in units of $z_0$. }
\label{fig:psiplots}
\end{center}

\vglue -5 mm
\end{figure}

%%%%%%%%%%%%%%%%%%%%%%%%%

Figure \ref{fig:psiplots} shows wave functions of the three lightest vector mesons. By our choice of boundary conditions hadrons wave functions localized closer to IR brane than to UV brane.

The presence of $J_0(\tilde{q} z_0)$ on the denominator in Eq.(\ref{profileV}) indicates the existence of poles for timelike $q$. More precisely, the profile function can be written as a sum over mesons poles 
\be
{\cal V}^a(q^2,z)=\sum_n \frac{-g_5 F^a_n \psi^a_n(z)}{q^2-{M^a_n}^2}\,,
\ee
where $F^a_n=|\partial_z\psi_n(\varepsilon)/(g_5\varepsilon)|$. From Eq.(\ref{2pointfV}), $F^a_n$ can be identified as  the decay constant of the $n$-th KK vector meson, 
\be
\left<0\left|{J^a_\mu}_\perp(0)\right| V_n^b(q,\lambda)\right>=F^a_n \varepsilon_\mu(q,\lambda) \delta^{ab}.
\ee

One can substitute the above expansion of the profile function into (\ref{2pointfV}) and obtain self energy function as a sum over narrow mesons poles. A well known signature of large $N_c$ QCD, which is intrinsic to the AdS/QCD correspondence. 
%%%%%%%%%%%%%%%%%%%%%%%%%%%%%%%%%%%%%%%%%%%%%%%%%%%%%%%%%%%%%%%
\subsection{Axial sector}
%%%%%%%%%%%%%%%%%%%%%%%%%%%%%%%%%%%%%%%%%%%%%%%%%%%%%%%%%%%%%%%%

Many of our derivations for the axial sector resemble corresponding derivations for the vector sector.
For example, the equation satisfied by the transverse part of the axial-vector field is analogous to Eq. (\ref{profiletrans}), with $\alpha^a$ replaced by $\beta^a(z)$.

The profile functions of the longitudinal part of the axial-vector field and the $\pi$ field satisfy the following equations 
\begin{align}
-q^2\partial_z \phi^a(q^2,z)&+\beta^a(z)\partial_z \pi^a(q^2,z)=0\,,
				\label{phipi_profile1}	\\[1.25ex]
\partial_z \left(\frac{1}{z}\partial_z\phi^a(q^2,z) \right)  &-\frac{\beta^a(z)}{z}\left(\phi^a(q^2,z)-\pi^a(q^2,z)\right)=0\,, 					\label{phipi_profile2}
\end{align}
where longitudinal part of the axial-vector field denoted by $A^a_{\mu\parallel}(x,z)=\partial_\mu \phi^a(x,z)$. The  boundary conditions are $\phi^a(q^2,\varepsilon)=0$, $\pi^a(q^2,\varepsilon)=-1$, and $\partial_z\phi^a(q^2,z_0)=\partial_z\pi^a(q^2,z_0)=0$. Note that these equations as well as boundary conditions are analogous to the longitudinal part of the vector field. 

In order to solve the coupled equations, one can combine Eq. (\ref{phipi_profile1}) and (\ref{phipi_profile2}) into a second order differential equation, defining $y^a(q^2,z)=\partial_z \phi^a(q^2,z)/z$, to obtain~\cite{Kwee:2007dd}
\be
\partial_z\left(\frac{z}{\beta^a(z)}\partial_z y^a(q^2,z)\right)+z\left(\frac{q^2}{\beta^a(z)}-1\right)y^a(q^2,z)=0\,. \label{eom}
\ee
In this notation, the boundary condition at the IR limit is given by $y^a(q^2,z_0)=0$ which is nothing but $\partial_z\phi^a(q^2,z_0)=0$. At the UV boundary,   Eq.~(\ref{phipi_profile2}) and boundary conditions of $\pi$ and $\phi$, give $\varepsilon\partial_z y^a(q^2,\varepsilon)/\beta(\varepsilon)=1$. Near the UV cut-off the profile function can be written as
\be
y^a(q^2,\varepsilon)=\beta^a(\varepsilon) \ln(q\varepsilon) + c_2 (q \varepsilon)^2 \ln(q\varepsilon) +\dots \,.
\ee 
Notice that although the above solution blows up logarithmically at UV, the profile functions  $\phi(q^2,z)$ as well as $\pi(q^2,z)$ do not, because of a multiplication by $z$ in their definition. 

%%%%%%%%%%%%%%%%%%%%%%%%
\begin{figure}[t]
\begin{center}
%\vglue -10 mm
\includegraphics[width =2.8in]{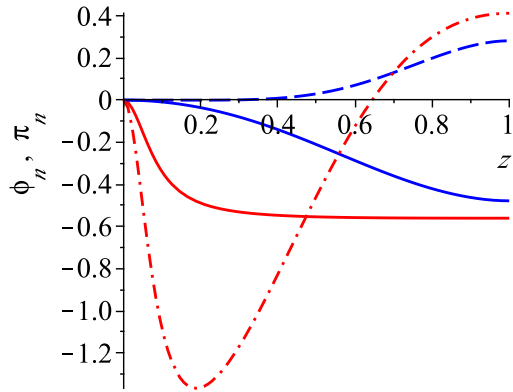}
%\vglue -5 mm
\caption{Plot of $\pi^a_1$ (lower solid curve, in red), and $\phi^a_1$ (upper solid curve, in blue) which correspond to $m^a_1=139.6$ MeV, and  $\pi^a_2$ (dash-dot red curve), and $\phi^a_2$ (dashed blue curve) which correspond to $m^a_2=1892.3$ MeV, for $a=1,2,3$, with $z$ in units of $z_0$. The units of $\phi_n^a$ and $\pi_n^a$ are $z_0^{-1}$. }
\label{fig:phipiplots}
\end{center}

\vglue -5 mm
\end{figure}
%%%%%%%%%%%%%%%%%%%%%%%%%

Pseudoscalar hadrons in the axial sector are pions, kaons and etas. Explicitly, $\pi^a$ field can be written as follows 
\be
\pi^a t^a=\frac{1}{\sqrt{2}}\left(\begin{array}{ccc}
                  \frac{\pi^0}{\sqrt{2}}+\frac{\eta^0}{\sqrt{6}} &   \pi^+      & K^+\\
                    \pi^-     &  -\frac{\pi^0}{\sqrt{2}}+\frac{\eta^0}{\sqrt{6}}  & K^0\\
                    K^-     &  \bar{K}^0       &-2\frac{\eta^0}{\sqrt{6}}
                  \end{array}
	\right)\,.
\ee
The corresponding normalizable modes, denoted by $y^a_n(z)$,  satisfy Eq. (\ref{eom}). Their  eigenvalues, $q^2={m^a_n}^2$, are the squared mass of the corresponding hadrons.  As in the vector sector, the axial sector allows not a single mode but an infinite tower of KK modes.  These modes satisfy $y^a_n(z_0)=0$ at the IR boundary, consistent with $\partial_z\phi^a_n(z_0)=\partial_z \pi^a_n(z_0)=0$, and $\varepsilon \partial_z y^a_n(\varepsilon)/\beta^a(\varepsilon)=0$ at the UV brane, consistent with $\phi^a_n(\varepsilon)=\pi^a(\varepsilon)=0$. 
Near the UV boundary, the normalizable mode behaves like
\be
y_n(z)=a_0+a_2 z^2+\ldots\,,
\ee 
or equivalently $\phi_n\sim a_0 z^2/2$ and $\pi_n\sim m_n^2 a_0 z^2/(2\beta^a(\varepsilon))$. The coeffiecient $a_0$ is determined by the orthonormality condition
\be
\int_\varepsilon^{z_0} dz\, \frac{z}{\beta^a(z)} y^a_n(z) y^a_m(z)=\frac{\delta_{nm}}{{m^a_n}^2}\,. \label{ynorm}
\ee
These normalizable wave functions can be solved numerically.  Using this normalization, the dimension of the normalizable modes is different from the dimension of the profile function.  We use it because it is well behaved for the ground state in the chiral limit (despite the $1/{m_n^a}^2$ on the right-hand-side).  A plot of $\pi_n$ and $\phi_n$, for $n=1$ and $n=2$, is shown in Fig.~\ref{fig:phipiplots}.

Let us derive how  the profile function, $y^a(q,z)$, can be written as sum over normalizable modes, $y^a_n(z)$. First, we write $y^a(q,z)=\sum c^a_n(q^2) y^a_n(z)$. Multiplying the left and the right hand side of the equation by $z(q^2-{m^a_n}^2)y_m(z)/\beta^a(z)$, then integrating over $z$, one obtains
\begin{align}
&c^a_m(q^2) \frac{\left(q^2-{m^a_m}^2\right)}{{m^a_m}^2}= -\frac{z}{\beta^a(z)} y^a_m(z) \partial_z y^a(q^2,z) \bigg|^{z_0}_\epsilon
				\nonumber\\
&  \hskip 15 mm
+\frac{z}{\beta^a(z)}y^a(q^2,z)\partial_z y^a_m(z) \bigg|^{z_0}_\epsilon\,,
\end{align}
after integration by parts and imposing the equation of motion (\ref{eom}). The second term and the upper limit of the first term vanish by the boundary conditions of $y^a(q,z)$ and $y^a_n(z)$. Hence, ignoring the non-pole terms, the profile functions can be written as
\be
y^a(q^2,\varepsilon)=\sum_n\frac{{m^a_n}^2y^a_n(\varepsilon) y^a_n(z)}{q^2-{m^a_n}^2}\,,
\ee
which can be integrated to obtain
\ba
\phi^a(q^2,z)&=&\sum_n\frac{- g_5 {m^a_n}^2 f^a_n\phi^a_n(z)}{q^2-{m^a_n}^2}\,,\label{profilefsum}
			\nonumber \\
\pi^a(q^2,z)&=&\sum_n\frac{- g_5 {m^a_n}^2 f^a_n\pi^a_n(z)}{q^2-{m^a_n}^2}\,,			
\ea
where $f^a_n=-\partial_z \phi^a_n(\varepsilon)/(g_5 \varepsilon)$.

Axial current-current correlators are  analogous to Eq.~(\ref{2pointfV}),
\begin{align}
i\int_x e^{iqx} \big< 0 \big| \mathcal{T} 
	J^{\mu a}_{A\perp}(x)J^{\nu b}_{A\perp}(0) \big| 0 \big>
		&=-P^{\mu\nu}_T \delta^{ab}
\frac{\partial_z {\cal A}^a(q^2,\varepsilon)}{g_5^2 \varepsilon}\,,							\nonumber\\
i\int_x e^{iqx}\big< 0 \big| \mathcal{T} 
	J^{\mu a}_{A\parallel}(x)J^{\nu b}_{A\parallel}(0) \big| 0 \big>
		&=-P^{\mu\nu}_L \delta^{ab}
\frac{\partial_z \phi^a(q^2,\varepsilon)}{g_5^2 \varepsilon}\,,
\end{align}
from which, one can identify $f^a_n$ as the decay constant,
\be
\left<0\left|J^a_{A\mu\parallel}(0)\right|\pi^b(q)\right>=i f^a_n q_\mu \delta^{ab}.
\ee

 The value $M^\rho_1=775.5 $ MeV gives $z_0= (322.5\, {\rm MeV})^{-1} $.  Parameters $m_q$ and $\sigma_q$ can be determined by fitting $m^a_1$ and $f^a_1$, for $a=1,2,3$, with the pion's mass and pion's decay constant respectively. Given experimental data for $m_\pi=139.6$ MeV and $f_\pi=92.4$ MeV, we obtain $\sigma_q= (\sqrt{3}/(2\pi))\,(328.3 {\rm\ MeV})^3 = (213.7 {\rm\ MeV})^3$ and $m_q=(2\pi/\sqrt{3}) \, 2.29 {\rm\ MeV} = 8.31 {\rm\ MeV}$.  In the $\sigma_q=\sigma_s$ limit, fitting $m^a_1$, for $a=4,5,6,7$,  with the kaon's mass, $m_K=495.7$ MeV, gives $m_s=(2\pi/\sqrt{3}) \, 51.96 {\rm\ MeV} = 188.5 {\rm\ MeV}$ (model A). A global fit to fifteen observables  (model B) yields $z_0 = (328.0 {\rm\ MeV})^{-1}$, $m_q = (2\pi/\sqrt{3}) \, 2.16 {\rm\ MeV} = 7.84 {\rm\ MeV}$, $\sigma_q=(\sqrt{3}/(2\pi))\,(312.2 {\rm\ MeV})^3= (203.2 {\rm\ MeV})^3 $, $m_s = (2\pi/\sqrt{3}) \, 56.81 {\rm\ MeV} = 206.1 {\rm\ MeV}$,  and $ \sigma_s = (\sqrt{3}/(2\pi))\,(322.8  {\rm\ MeV})^3 = (210.1 {\rm\ MeV})^3$. The fifteen obervables include eleven observable in Table~\ref{table1} and the additional four observables are $f_{+}(0)$, $\lambda'_+$, $\lambda''_+$  and $\lambda_0$, from the kaon to pion transition form factor discussed in Sec.\ref{sec:numerical}. The quark masses given here include the normalization parameter suggested in~\cite{DaRold:2005vr,Cherman:2008eh}.  The AdS/QCD quark masses are renormalization scale independent.   In QCD the quark masses do evolve with renormalization scale.  We should compare our masses with experimental QCD values of the quark masses at a low renormalization scale, say 1 GeV or perhaps a bit below.  The quark masses quoted by the particle data group~\cite{Amsler:2008zzb} evolved to 1 GeV using their prescriptions are in the range 3.4--7 MeV for $m_q$ and 95--175 MeV for $m_s$.      Predictions of the model for masses and decay constants using terms up to second order expansion in the fields of the 5D action are summarized in Table~\ref{table1}.   These may be compared to results in~\cite{Shock:2006qy,Katz:2007tf,Erlich:2008en}.

\begin{table}[t]
\caption{Masses and decay constants.  Model A is a four parameter fit to four observables as indicated in the Table, and maintains $\sigma_s = \sigma_q$.  Model B is a five parameter fit to 15 observables (11 from this Table and 4 from the kaon to pion transition form factors discussed in the next section) with $\sigma_s \ne \sigma_q$.  The values of the parameters are given in the text.
%These were obtained using parameters $z_0 = (322.5 {\rm\ MeV})^{-1}$, $m_q = (2\pi/\sqrt{3}) \, 2.29 {\rm\ MeV} = 8.31 {\rm\ MeV}$, $\sigma_q=\sigma_s = (\sqrt{3}/(2\pi))\,(328.3 {\rm\ MeV})^3 = (213.7 {\rm\ MeV})^3$, for model~A, and
%$z_0 = (328.0 {\rm\ MeV})^{-1}$, $m_q = (2\pi/\sqrt{3}) \, 2.16 {\rm\ MeV} = 7.84 {\rm\ MeV}$, $\sigma_q=(\sqrt{3}/(2\pi))\,(312.2 {\rm\ MeV})^3= (203.2 {\rm\ MeV})^3 $, $m_s = (2\pi/\sqrt{3}) \, 56.81 {\rm\ MeV} = 206.1 {\rm\ MeV}$,  $ \sigma_s = (\sqrt{3}/(2\pi))\,(322.8  {\rm\ MeV})^3 = (210.1 {\rm\ MeV})^3$, for model~B.
}
\renewcommand{\tabcolsep}{0.3cm}
\begin{tabular}{|c|c|c|c|}
\hline
\hline
Observable & Model A& Model B & Measured \\
		&  ($\sigma_s = \sigma_q$) &	($\sigma_s \ne \sigma_q$)	&	\\
                       & (MeV)      & (MeV) & (MeV)\\
\hline
\hline
$m_\pi$  &  (fit)  &  134.3 &  $139.6$  \\
$f_\pi$  &   (fit)   &  86.6 & $92.4$   \\
$m_K$  &    (fit)   &  $513.8$ & $495.7$  \\
$f_K$ & $104$ & $101$ & $113 \pm 1.4$\\
\hline
$m_{K^*_0}$&$791 $& $697$ & $672$\\
$f_{K^*_0}$ & $28.$ & $36$ & $$\\
\hline
$m_\rho$  &  (fit)  & 788.8 & $775.5$  \\
$F_{\rho}^{1/2}$ & $329$ & $335$ & $345\pm 8$\\
$m_{K^*}$&$791 $& $821$ & $893.8$\\
$F_{K^*}^{1/2}$ & $329$ & $337$ & $$\\
\hline
$m_{a_1}$&$1366 $&$1267 $&$1230\pm 40$\\
$F_{a_1}^{1/2}$&$489 $&$453 $&$433\pm 13$\\
$m_{K_1}$&$1458 $&$1402 $&$1272\pm 7$\\
$F_{K_1}^{1/2}$&$511 $&$488 $&$ $\\
\hline
\hline
\end{tabular}
\label{table1}
\end{table}

%%%%%%%%%%%
\subsection{Massless pion limit}

The AdS/QCD model has consequences of chiral symmetries, such as the Gell-Mann--Oakes--Renner relation (GOR), as shown in \cite{Erlich:2005qh}. Here, we will present a slightly different derivation, starting from the normalization condition, Eq.(\ref{ynorm}). As noted in \cite{Erlich:2005qh}, the weight function $z/\beta(z)$ has significant support only for $z$ close to $z_c=\sqrt{m_q/3\sigma}$, hence, the normalizable wave function $y_n$ can be evaluated at $z\sim \varepsilon$ and moved outside the integral.  Away from $z_c$, the weight function decays very fast, hence the upper limit integral can be replaced by infinity. Noting that from just after Eq.~(\ref{profilefsum}), $y_n(\varepsilon)=-g_5 f_\pi$, we obtain 
\be
g_5^2 {f^a_1}^2 {m^a_1}^2 \int_0^\infty dz \frac{z}{\beta^a(z)}=1.
\ee 
The GOR relations  immediately follow, $f_\pi^2 m_\pi^2=2 m_q \sigma_q$ for $a=1,2,3$ and $f_K^2 m_K^2=(m_q+m_s)(\sigma_q+\sigma_s)/2$ for $a=4,5,6,7$. However, for the kaon case, our results deviate by over 30 percent from the GOR relation.

As $m_\pi$ approaches zero, the GOR relation becomes exact. Fixing $f_\pi$ to experimental data, parameter $\sigma_q$ approaches $(331.6\,{\rm MeV})^3$. In this limit, the $\pi_1(z)$ normalizable wave function becomes constant, $\pi_1(z)=-1/(g_5 f_\pi)$, throughout the region of interest with a step function-like jump near the UV boundary. The wave function of the lightest mode can be solved analytically in terms of modifed Bessel function. Defining $\eta=g_5\sigma/3$, one obtains 
\be
y_1=N z^2\bigg(-I_{-\frac{2}{3}}\left(\eta z^3\right)+\frac{I_{-\frac{2}{3}}\left(\eta z_0^3\right)}{I_{\frac{2}{3}}\left(\eta z_0^3\right)}I_{\frac{2}{3}}\left(\eta z^3\right)\bigg)\,,
\ee
where,
\be
N^2=g_5^2 \sigma^2 \Gamma(2/3)\Gamma(1/3)\frac{I_{\frac{2}{3}}\left(\eta z_0^3\right)}{ I_{-\frac{2}{3}}\left(\eta z_0^3\right)}\,.
\ee
Evaluating $y_1$ at UV boundary, we obtain  an equation relating $f_\pi$ and $\sigma$ in the chiral limit which is in perfect agreement with previous result \cite{Grigoryan:2007wn}. One should notice that $n>1$ KK pions still present in the chiral limit.

%%%%%%%%%%%%%%%%%%%%%%%%%%%%%%%%
%
\section{Three point functions and form factors} \label{sec:threepoint}
%
%%%%%%%%%%%%%%%%%%%%%%%%%%%%%%%%

The transition form factors for $K_{\ell 3}$ decay is defined from~\cite{Amsler:2008zzb,Antonelli:2008jg}
\begin{align}
&	\langle \pi^-(p) |
	J_\mu^{(|\Delta S| = 1)} | K^0(k) \rangle
				\nonumber \\
& \hskip 10 mm	= f_+(q^2) (k+p)_\mu + f_-(q^2) (k-p)_\mu\,,
%				\nonumber \\
%& \hskip 10 mm =	f_+(q^2) \left( k_\mu + p_\mu
%				- \frac{m_K^2 - m_\pi^2}{q^2} q_\mu \right)
%				\nonumber \\
%& \hskip 20 mm	+ f_0(q^2) \frac{m_K^2 - m_\pi^2}{q^2} q_\mu
\end{align}
with $q = k - p$.  By isospin, they could equally well be defined using the $K^+ \!\!\to \pi^0$ transition.  Only the vector part of the current contributes.  Further let
\be
f_0(q^2) = f_+(q^2) + \frac{q^2}{m_K^2 - m_\pi^2} f_-(q^2)\,,
\ee
so that $f_+$ and $f_0$ come from the transverse and longitudinal parts, respectively, of $J_\mu^{(|\Delta S| = 1)}$.  Unless $f_-(q^2)$ diverges as $q^2 \to 0$, one has $f_+(0) = f_0(0)$.  One may also write
\be
J_\mu^{(|\Delta S| = 1)} = J_\mu^4 +i J_\mu^5\,,
\ee
to show the SU(3) flavor indices.

In AdS/CFT or AdS/QCD, the three-point functions involving three currents are obtained by functionally differentiating the 5D action with respect to their sources, which are taken to be boundary values of the 5D fields that have the correct quantum numbers~\cite{Witten:1998qj,Gubser:1998bc,Grigoryan:2007vg,Abidin:2008ku}.  To wit,
\begin{align}
\label{3pf}
&\langle 0 |T J_{A\parallel}^{\alpha a} (x) J_{\perp}^{\mu b}(y) 
	J_{A\parallel}^{\beta b}(w)	| 0 \rangle
=\frac{(i/i^3)\ \delta S(V\pi\pi) \qquad } 
	{\delta A_{\parallel\alpha}^{0a}(x) \,
	\delta V_{\perp\mu}^{0b}(y) \, \delta A_{\parallel\beta}^{0c}(w)
	}
\end{align}
where $S(V\pi\pi)$ is the relevant part of the 5D action evaluated using classical fields that solve the equations of motion with $z=0$ boundary values $A_{\parallel\alpha}^{0a}(x)$ or $V_{\perp\mu}^{0b}(y)$. 

Matrix elements of the current are obtained from the three-point functions using~\cite{Grigoryan:2007vg,Abidin:2008ku},
\begin{align}
& \langle \pi_n^a(p) | J_{\perp}^{\mu b}(0) | \pi_k^c(k) \rangle =
				\nonumber  \\[1ex]
& \hskip 2mm  \lim_{ \stackrel{\scriptstyle{p^2\to {m_n^a}^2}}{k^2\to {m_k^c}^2}}
%\lim_{  {p^2\to {m_n^a}^2}, \, {k^2\to {m_k^c}^2} }
	\ 	\frac{p_\alpha k_\beta}{p^2 k^2}
  \frac{ (p^2 - {m_n^a}^2) (k^2 - {m_k^c}^2)}{f_{\pi_n^a} f_{\pi_k^c}} 
  				\nonumber  \\
& \hskip 4mm  \times  	\int d^4x\, d^4w  e^{ipx - ikw}   \ 
 	\langle 0 |T J_{A\parallel}^{\alpha a} (x) J_{\perp}^{\mu b}(0) 
		J_{A\parallel}^{\beta c}(w)	| 0 \rangle	\,,
\end{align}
from which we obtain the form factor $f_+$.  

A similar expression for the longitudinal part of the current $J_{\parallel}^{\mu b}(0)$ using $V_{\parallel\mu}^{0b}(y)$, allowing us to obtain the scalar form factor $f_0$.

The relevant part of the action receives contributions both from the gauge terms and the chiral terms.  Keeping only terms that have one vector field and two pion fields (either $\phi^a(x,z)$ or $\pi^a(x,z)$) one obtains
\ba
\label{vpipiaction}
S(V\pi\pi) &=& \int d^5x \bigg\{ \frac{1}{2g_5^2 z}
	f^{abc}
				\nonumber \\[1ex]
&& \hskip 9 mm	\times \ 
	\left( \partial^\mu \phi^a V_{\mu\nu}^b 
		\partial^\nu \phi^c 
		+ 2 \partial_z \partial_\nu \phi^a V_z^b 
			\partial^\nu \phi^c \right)
				\nonumber \\
&+& \frac{1}{z^3} \bigg[ g^{abc} \left( 
	\partial^\mu \pi^a - \partial^\mu \phi^a \right)
		V_\mu^b \pi^c
				\\ \nonumber
&&  \hskip 6 mm
		- \ h^{abc} \left(  \frac{1}{2} 
			\partial^\mu \left( \pi^a \pi^c \right)
			- \partial^\mu \phi^a \pi^c \right) V_\mu^b \bigg]
				\\ \nonumber
&-& \frac{1}{z^3} \left[ g^{abc}  \partial_z \pi^a V_z^b \pi^c 
	- h^{abc}  \frac{1}{2} \partial_z \left( \pi^a \pi^c \right) 
		V_z^b \right]		\bigg\}
\ea

The $f^{abc}$ terms come from the gauge part of the original action, and the other terms come from the chiral part.  We have defined
\begin{align}
g^{abc} &= -2i {\rm \,Tr\,} \left\{t^a,X_0 \right\}
	\left[ t^b, \left\{ t^c, X_0 \right\} \right]
			\nonumber \\
h^{abc} &= -2i {\rm \,Tr\,} \left[t^b,X_0 \right]
	\left\{ t^a, \left\{ t^c, X_0 \right\} \right\}
\end{align}
If none of $a$, $b$, or $c$ is equal to ``$8$'', these become
\begin{align}
g^{abc} &= f^{abc} v_a v_c	\,,
			\nonumber \\
h^{abc} &= f^{abc} (v_c - v_a) v_c	\,,
\end{align}
where for $X_0 = \frac{1}{2} c_0 + c_8 t^8$, 
\ba
v_a = c_0 + c_8 d^{aa8} =
	\left\{	\begin{array}{cl}
			v_q \,, & a=1,2,3 \\
			\frac{1}{2}\left( v_q + v_s \right) \,, & a=4,5,6,7 \,.
			\end{array}
	\right.
\ea

The derivatives indicated in Eq.~(\ref{3pf}) are facilitated by going to Fourier transform space and using the relations~\cite{Grigoryan:2007wn,Abidin:2008hn},
\begin{align}
\phi^a(p,z) &= \phi^a(p^2,z) \phi^{0a}(p) 
	= \phi^a(p^2,z) \frac{i p^\alpha}{p^2}
	A_{\parallel\alpha}^{0a}(p)	\,,
			\nonumber \\
\pi^a(p,z) &= \pi^a(p^2,z) \frac{i p^\alpha}{p^2}
	A_{\parallel\alpha}^{0a}(p)	\,,
			\nonumber \\
V_{\perp \mu}^b(q,z) &= {\cal V}^b(q^2,z)\ 
	V_{\perp \mu}^{0b}(q)	\,,
			\nonumber \\
V_{\parallel \mu}^b(q,z) %&= 
	%{\cal V}^b(q^2,z)\ V_{\parallel \mu}^{0b}(q)
			%\nonumber \\
	&= \big( \tilde \phi^b(q^2,z) - \tilde \pi^b(q^2,z) \big)
		\ V_{\parallel \mu}^{0b}(q)	\,,
			\nonumber \\
V_z^b(q,z) &= - \partial_z \tilde\pi^b(q^2,z) 
	\ \frac{i q^\alpha}{q^2}  V_{\parallel \mu}^{0b}(q)	\,,
			\nonumber \\
\partial^\mu &\to -i \left( {\rm relevant\ momentum} \right)^\mu\,.
\end{align}

\noindent  With experience, one can use the above translation dictionary to obtain form factor results quite quickly.  Incidentally, in the limit of having the same quark condensate parameter $\sigma$ for all flavors of quarks, one can show that the bulk-to-boundary propagator $\mathcal{V}^b(q^2,z)$ for the transverse case is identical to 
$\tilde \phi^b(q^2,z) - \tilde \pi^b(q^2,z)$.

For the transverse form factor, the $V_z$ terms in the action, Eq.~(\ref{vpipiaction}), do not contribute.   One obtains,
\begin{align}
& f_+(q^2) =  \int_0^{z_0} dz \, {\cal V}^4(q^2,z) 
\,	 \bigg\{ \frac{1}{z} \partial_z\phi^1(z) \partial_z\phi^7(z)
		\\ \nonumber
&	\hskip 5 mm  + \frac{g_5^2}{2z^3}      v_q (v_q + v_s)  
	\left( \phi^1(z) - \pi^1(z) \right)
	\left( \phi^7(z) - \pi^7(z) \right)
	\bigg\} 	,
\end{align}
where $\phi^a$ and $\pi^a$ are now ground state normalizable eigenmodes, with subscript ``1'' tacit.  The superscripts on $\phi^a$, $\pi^a$, and $\mathcal{V}^b$ are the flavor indices for quantities with pion or kaon quantum numbers.  We are working in the isospin conserving limit, so that the $\phi^a$ are the same for $a = 1,2,3$ and again the same for the set $a=4,5,6,7$, and similarly for $\pi^a$ and $\mathcal{V}^b$.

As a check, in the equal mass limit, $v_s=v_q=v$, the transverse $K_{\ell 3}$ form factor should by SU(3) symmetry be the same as the electromagnetic form factor.  One obtains in this limit
\begin{align}
& f_+(q^2) \stackrel{ \rm SU(3)\ limit}{=} \int_0^{z_0} dz \, {\cal V}(q^2,z)
		\nonumber \\
&	\hskip 0 mm \times \ 
	 \bigg\{  \frac{1}{z} (\partial_z \phi(z))^2
	+ \frac{g_5^2 v^2(z)}{z^3} \,  \left( \phi(z)- \pi(z) \right)^2
	\bigg\} 			,
\end{align}
which indeed is the same as $F_\pi(q^2)$ as found in Eq.~(3.5) in~\cite{Kwee:2007dd} or to Eq.~(38) in~\cite{Grigoryan:2007wn}, allowing for the fact that those authors wrote the results using the profile functions and the massless pion limit, whereas we used the normalizable eigensolutions~\cite{Abidin:2008ku} and nonzero mass.

The longitudinal form factor is
\begin{widetext}
\begin{align}
&	f_0(q^2) =
\int_0^{z_0}  dz   \Bigg(  \big( \tilde \phi^4(q^2,z) - \tilde \pi^4(q^2,z) \big)
	\Bigg\{   
	\frac{1}{z} \partial_z \phi^1 \partial_z \phi^7
	+ \frac{g_5^2 v_q ( v_q + v_s ) }{2z^3}(\phi^1-\pi^1)(\phi^7-\pi^7)
	+ \frac{q^2}{2 z} \phi^1 \phi^7 
				\nonumber \\
& \hskip 5 mm
+ \frac{ g_5^2 q^2}{8z^3 (m_K^2-m_\pi^2)} 
	\bigg[  ( v_s-v_q)(3v_q+v_s) (\phi^1-\pi^1)(\phi^7-\pi^7)
	- 4 v_qv_s \phi^1 (\phi^7-\pi^7)
	+ ( v_q + v_s)(3v_q - v_s) (\phi^1-\pi^1) \phi^7 \bigg] \Bigg\}
				\nonumber \\[1ex]
& + \frac {\partial_z \tilde\pi^4(q^2,z)} {(m_K^2-m_\pi^2)}
 \times \bigg\{    \frac{m_k^2 + m_\pi^2 -q^2 }{ 2 z }
    \left( \partial_z\phi^1 \,\phi^7 - \phi^1 \partial_z\phi^7 \right)
     				\nonumber \\[1ex]
& \hskip 5 mm
	+ \frac{ g_5^2 ( v_s-v_q) (3v_q + v_s) }{ 8z^3 } \partial_z (\pi^1 \pi^7)
	+ \frac{ g_5^2 v_q ( v_q+v_s) }{2 z^3}
	\left( \pi^1 \partial_z\pi^7 - \partial_z\pi^1 \,\pi^7 \right)
	- \frac{m_K^2 - m_\pi^2 }{ 2z }  
	\frac{\partial_z \alpha^4(z) }{\alpha^4(z)} \, \phi^1 \phi^7
	\bigg\}
	\Bigg)			.
\end{align}
\end{widetext}

\noindent  The identity $f_0(0) = f_+(0)$ is apparent after noting that 
$\partial_z\tilde\pi^b(0,z) = 0$ and considering the $q^2=0$ normalizations of the profile functions.

%%%%%%%%%%%%%%%%%%%%%%%%%%%%%%%%
%
\section{Numerical Results for ${K_{\ell 3}}$ Form Factors} \label{sec:numerical}
%
%%%%%%%%%%%%%%%%%%%%%%%%%%%%%%%%

\begin{figure}[htbp]
\includegraphics[width = 3.35 in]{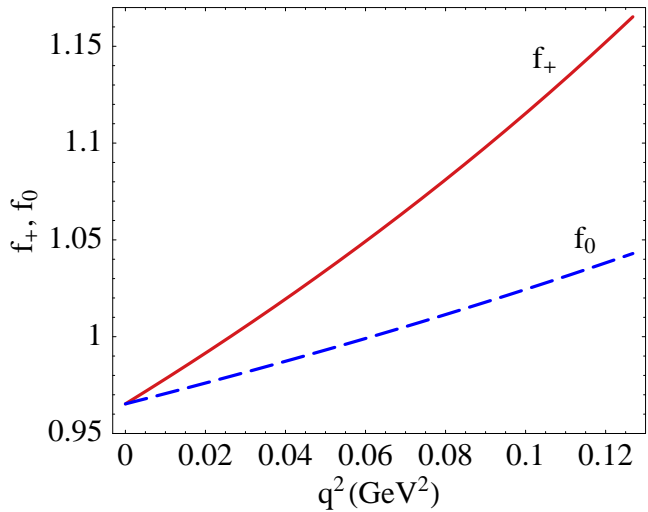}
\caption{The $K_{\ell 3}$ form factors $f_+$ (solid red line) and $f_0$ (dashed blue line) plotted vs. $q^2$ over the physical range pertinent to $K \to \pi e \nu$.  The plot is based on the  ``Model A'' parameters, where $\sigma_s=\sigma_q$.}
\label{fig:kell3ff}
\end{figure}

We obtain numerical solutions for the bulk-to-boundary propagators and the normalizable eigenfunctions in the massive quark case using Mathematica or Maple, and then use the numerical solutions to obtain $f_+(q^2)$ and $f_0(q^2)$.  We present a plot of the results in Fig.~\ref{fig:kell3ff}.  Of particular interest for comparison to data~\cite{Amsler:2008zzb,Antonelli:2008jg} and to chiral perturbation theory~\cite{Leutwyler:1984je,Bijnens:2003uy,Cirigliano:2005xn,Jamin:2006tj} or lattice gauge theory~\cite{Becirevic:2004ya,Dawson:2006qc,Boyle:2007qe,Flynn:2008hd,Lubicz:2009ht} are the values of $f_+(0)$, the slopes of $f_+$ and $f_0$, and the curvature of $f_+$.

%%%%%%%%%%%%%%%%%%%%%%%%

%%%%%%%%%%%%%%%%%%%%%%%%

Our results from models A ($\sigma_s = \sigma_q$) and B ($\sigma_s$ independently fit), as well as the results from lattice gauge theory, chiral perturbation theory, and experiment are listed in Table~\ref{table2}.

%%%%%%%%%%%%%%%%%%%%%%%

\begin{table*}[t]
\caption{Results from our models, compared to lattice gauge theory, chiral perturbation theory, and experimental data.}
\begin{center}
\renewcommand{\tabcolsep}{0.2 cm}
\begin{tabular}{|c|c|c|c|c|c|}
\hline
\hline
Observable & Model A & Model B & Lattice 
					& $\chi$PT & Data \cite{Antonelli:2008jg} \\
\hline
\hline
$f_+(0)$ & 0.965 & 0.936 	& 0.968(11) \cite{Dawson:2006qc} 
						& 0.961(8) \cite{Leutwyler:1984je} & \\
		&            &		& 0.9742(41) \cite{Flynn:2008hd} 
						& 0.978(10) \cite{Bijnens:2003uy} & \\
		&            &		& 0.9560(84) \cite{Lubicz:2009ht} 
						& 0.984(12) \cite{Cirigliano:2005xn} & \\
		&		&	&	& 0.974(11) \cite{Jamin:2006tj} & \\
\hline 
$\lambda'_+$ & 0.0249 & 0.0227 & 0.0237(23)(21) \cite{Lubicz:2009ht} 
			& & 0.0249(11)			\\
\hline
$\lambda''_+$ & 0.0021 & 0.0016 & & & 0.0016(5) \\
\hline
$\lambda_0$ & 0.0123 & 0.0140 & 0.0128(22)(45) \cite{Lubicz:2009ht} 
			& & 0.0134(12) \\
\hline
\hline
\end{tabular}
\end{center}
\label{table2}
\end{table*}%
%%%%%%%%%%%%%%%%%%%%%%%

Experiments measure $f_+(0)$ times the Cabibbo-Kobayashi-Maskawa (CKM) matrix element $V_{us}$.  If the CKM matrix element is gotten from elsewhere, for example from the unitarity relation, then all the above values of $f_+(0)$ are in agreement with experimental data.  More usually, the calculations are taken to be accurate within the stated limits, and are used to extract the most precise available values of $|V_{us}|$ from the data.  

Experiments also measure the slope and curvature of the $K_{\ell 3}$ form factors.  For $f_+$, both the slope and curvature can be fit, and are parameterized as~\cite{Amsler:2008zzb}
\be
f_+(q^2) = f_+(0) \left( 1 + \lambda'_+ \left(q^2/m_{\pi^+}^2 \right)
	+ \frac{1}{2} \lambda''_+ \left(q^2/m_{\pi^+}^2 \right)^2
	\right),
\ee
while for $f_0(q^2)$ there is a linear fit
\be
f_0(q^2) = f_0(0) \left( 1 + \lambda_0 \left(q^2/m_{\pi^+}^2 \right)
	 \right)
	\,.
\ee
Values for the parameters are given in the Table. For the experimental data in Table~\ref{table2}, we  took the numbers from the FlaviaNet Working Group on Kaon Decays~\cite{Antonelli:2008jg}.

Additionally,~\cite{Flynn:2008hd} quotes a result $f_-(0) = -0.113 (12)$.
The intercept $f_-(0)$ can be related to the slope parameters,
\be
f_-(0) = \frac{m_K^2 -m_\pi^2}{m_{\pi^+}^2} f_+(0) 
	\left( \lambda_0 - \lambda'_+ \right)	,
\ee
which leads to $f_-(0) = -0.141$ for model A and $f_-(0) = -0.110$ for model B  obtained here or $f_-(0) = -0.129 (18)$ using the FlaviaNet fits to experimental data for $\lambda_0$ and $\lambda'_+$.

%%%%%%%%%%%%%%%%%%%%%%%%%%%%%%%%
%
\section{Conclusion} \label{sec:end}
%
%%%%%%%%%%%%%%%%%%%%%%%%%%%%%%%%

We have extended the AdS/QCD model of Ref. \cite{Erlich:2005qh,DaRold:2005zs} to $SU(3)_L\times SU(3)_R$ model with a broken flavor symmetry. In order to introduce quarks with differing masses, we write the $X$ field with the exponentials of the pseudoscalar field split symmetrically about the classical expectation value $X_0$.  We find that neither the longitudinal part of the vector field nor $V_z$ can be gauged away for $a=4,5,6,7$.   If instead of using the symmetric form, one expands $X=\exp(i2\pi^a t^a) X_0 $, as can certainly be done when $X_0$ is a multiple of the identity, the longitudinal part of the vector field and $V_z$ will in general mix with the $\pi$ field, hence with the longitudinal part of the axial-vector field, and give a parity violating term in the Lagrangian.

We have done both a four parameter and a five parameter fit.  The four parameter fit constrains the condensate parameter to be flavor symmetric,  $\sigma_s = \sigma_q$, and the other four parameters are fit to the rho meson's mass, pion's mass, pion's decay constant and kaon's mass.   Predictions of the model for the non-dynamical properties of mesons such as masses and decay constants are within $20\%$ of the experimental data.   There is an infinite KK tower of pions present,  just as there is for vector and axial vector mesons.     For three-point functions, we have calculated kaon-to-pion transition form factors, $f_+$ and $f_0$ and obtain excellent agreement with experiment for the slope as well as for the curvature. We further find that the intercept, $f_+(0)$,  agrees very well with lattice and chiral perturbation theory calculations.

The five parameter fit allows $\sigma_s$ to vary from $\sigma_q$, and we performed a global fit to fifteen observables, including the intercept, slope, and curvature of the $K$ to $\pi$ transition form factors.   The results were again good, comparable to though somewhat improved as expected compared to the four parameter fit.  The best value of the strange condensate parameter was close to the value in the non-strange sector.

We could perhaps add that we found the intercept $f_+(0)$ to be somewhat touchy.  A drift of either $m_s$ or $\sigma_s$ away from the best values could lead to a significant decrease in its value.  On the technical side, in the chiral limit, $f_+(0)$ becomes normalized to unity.  This is because the profile functions 
${\cal V}^b(q^2,z)$ in the chiral limit are unity at $q^2 = 0$ for all $z$ and all $b$, so that $f_+(0)$ becomes just a wave function normalization integral.  However, when differing quark masses and differing condensate parameters are used, the profile function at $q^2=0$ is unity only for $z=0$ and can drift quite far from unity as $z$ approaches the IR cutoff, particularly if $\sigma_s$ gets far from $\sigma_q$.

A lack within the AdS/QCD framework is the absence of error estimates.  On the other hand, in the present context, it should be remembered that the form factor at $q^2 = 0$ is fixed by a normalization requirement in the equal quark mass limit, so what is really being calculated for $f_{+,0}(0)$ is the difference away from unity.  Here a 10--20\% error suffices match the error estimates quoted from other methods.

Possible extensions of present work include calculations of four-point  functions to obtain the $B_K$ parameter and $K_{\pi2}$ decay amplitudes~\cite{Hambye:2006av}, to consider isospin breaking in the context of the present model, and to use AdS/QCD as a method to study how differing quark mass and hence differing pion mass affects the calculated results, and compare the trends that are found to lattice gauge theory.  We hope to return to these topics in future work.

%%%%%%%%%%%%%%%%%%%%%%%%%%%%%%%

\begin{acknowledgments}

We thank Christopher Aubin, Will Detmold, and Josh Erlich for helpful comments, and the National Science Foundation for Grants numbered PHY-0555600 and PHY-0855618.

\end{acknowledgments}

%%%%%%%%%%%%%%%%%%%%%%%%%%%%%%%

\bibliography{adscft}

\end{document}